\documentclass[prl,floatfix,twocolumn,superscriptaddress,showpacs]{revtex4}
\usepackage{graphicx}
\begin{document}
\title{Resonance Widths in Open Microwave Cavities studied by Harmonic Inversion}
\author{U.~Kuhl}
\affiliation{Fachbereich Physik, Philipps-Universit\"{a}t Marburg, Renthof 5,
35032 Marburg, Germany}
\author{R.~H\"{o}hmann}
\affiliation{Fachbereich Physik, Philipps-Universit\"{a}t Marburg, Renthof 5,
35032 Marburg, Germany}
\author{J. Main}
\affiliation{Institut f\"{u}r Theoretische Physik 1, Universit\"{a}t Stuttgart,
70550 Stuttgart, Germany}
\author{H.-J.~St\"{o}ckmann}
\affiliation{Fachbereich Physik, Philipps-Universit\"{a}t Marburg, Renthof 5,
35032 Marburg, Germany}

\date{\today}

\begin{abstract}
From the measurement of a reflection spectrum of an open microwave cavity the poles of the scattering matrix in the complex plane have been determined. The resonances have been extracted by means of the harmonic inversion method. By this it became possible to resolve the resonances in a regime where the line widths exceed the mean level spacing up to a factor of 10, a value inaccessible in experiments up to now. The obtained experimental distributions of line widths were found to be in perfect agreement with predictions from random matrix theory when wall absorption and fluctuations caused by couplings to additional channels are considered.
\end{abstract}


\maketitle

Nowadays open wave-chaotic systems are intensely investigated experimentally and theoretically (for recent reviews see \cite{kuh05b,fyo05a,stoe99}). Especially effects of coupling, absorption, and decoherence are of interest. This aspect has been studied experimentally in microwave cavities \cite{men03a,kuh05a,hem05a,bar05b} and acoustics \cite{lob03a}, but also in semiconductor devices such as quantum dots \cite{aki97}. Till now mainly reflection and transmission amplitudes and phases of the spectra have been investigated, as it is extremely difficult to resolve the resonances in the regime of strong overlap. In the case of non-overlapping resonances, i.\,e.\ weak coupling and weak absorption, the width distributions are expected to obey a $\chi^2$ distribution \cite{fyo05a} which has been found experimentally in microwave cavities \cite{alt95a}. There is only one  experimental work studying resonance widths in the overlapping regime by Persson et al.\ \cite{per00}, where the effect of resonance trapping was investigated. In the latter work the centered time-delay analysis (CTDA) was used to extract the resonance widths.

Till now an experimental verification of theoretical and numerical investigations of the resonance width in dependence of the number of channels and their coupling strengths \cite{fyo97a,fyo97b,som99} had been missing. In the regime of strong resonance overlap the CTDA method does no longer work. In this regime the harmonic inversion (HI) method poses an alternative to extract resonances from a spectrum. The HI method, introduced by Wall and Neuhauser \cite{wal95} and  improved by Mandelshtam, Taylor, Main and others \cite{man97b,mai99}, allows to determine the unknown complex frequencies $\nu_n -i \gamma_n$ and amplitudes $a_n$ from a time series $C(t)=\sum_n a_n \exp[-2\pi i(\nu_n -i \gamma_n) t]$. The $\nu_n$, $\gamma_n$, and $a_n$ are obtained from a set of nonlinear equations which can be solved numerically stable. A description of the method for the extraction of resonance positions and widths can be found in Ref.~\onlinecite{wie08}. In this Letter, the method is applied, for the first time, to analyze experimental absorption spectra.

In the experiment a flat microwave cavity of the shape of a half Sinai billiard (length $a$=\,43\,cm, width $b$=23.7\,cm) has been used, with a half disc of diameter $d=12$\,cm attached to the long side. The reflection coefficient was measured in the range from 1 to 19.4\,GHz by means of a single wire antenna attached to the cavity. In this frequency range the cavity with a height of 7.8\,mm may be treated as two-dimensional. To improve statistics the half disc was moved along the wall in steps of 5\,mm to obtain 57 measurements. The complex $S$ matrix was measured with a vector network analyzer. Coupling strength of the antenna $T_a$ has been determined by the mean values of $S$ ($T_a=1-|\langle S \rangle|^2$) and the total absorption strength $T_t$ by the decay of the Fourier transformed autocorrelation function \cite{sch03a}.
For the evaluation of the width distribution it showed up to be necessary to split $T_t$ into two parts, one constant contribution  $T_{\rm const}$ and a fluctuating one $T_{\rm fluc}$, the latter one being mimicked by the coupling of $N_c$ fictitious absorbing channels to the billiard, $T_{\rm fluc}=N_c\cdot T_c$ \cite{sav06}.  The complete range from weak absorption and weak coupling ($T_t=0.75$, $T_a=0.116$, 4 to 5\,GHz) to strong absorption and nearly perfect coupling ($T_t=7.0$, $T_a=0.989$, 14.7 to 15.7\,GHz) has been covered. The antenna corresponds to a single channel since its radius $r=0.2$\,mm is much smaller than the wavelength. The same system had been considered already previously in Refs.~\cite{men03a,kuh05a}, where the reflection coefficient and the Poisson kernel had been investigated. The values for $T_t$ in Refs.~\cite{men03a,kuh05a} slightly differ from the ones given above, since  the fluctuating part of $T_t$ gives rise to a deviation from the exponential decay which has not been accounted for in the previous works.

\begin{figure}
\includegraphics[width=.8\columnwidth]{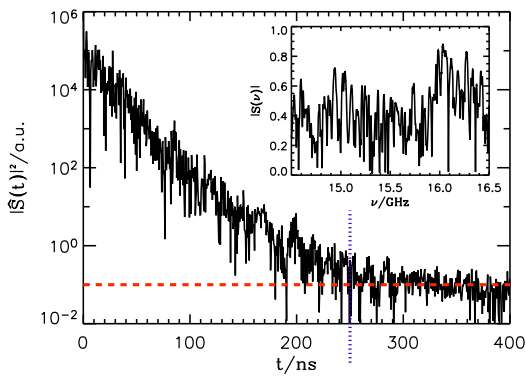}
\caption{\label{fig:Time} (Color online)
Modulus square $|\hat{S}(t)|^2$ of the Fourier transform of the part of the spectrum $S(\nu)$ ranging from 14.5 to 16.5\,GHz (see inset). The dashed horizontal line indicates the noise level of the Fourier transform. The dotted vertical line shows the cut-off time applied for the harmonic inversion. }
\end{figure}

An example of a spectrum in the strongly overlapping regime is shown in Fig.~\ref{fig:Time} as an inset. The spectrum is a superposition of Lorentzians,
\begin{equation}
  \label{eq:SumOfLorentzians}
  S(\nu)=1-\sum_n\frac{a_n}{\nu-\nu_n+i\gamma_n},
\end{equation}
For the application of the HI technique we first have to transform the spectrum to the time domain via a Fourier transform,
\begin{equation}
  \hat{S}(t)
  =\frac{1}{2\pi}\int\limits_{-\infty}^\infty e^{-2\pi i\nu t} S(\nu) {\rm{d}}\nu
  = \delta(t) - \sum_n a_n  e^{-2\pi i(\nu_n-i\gamma_n) t}
\end{equation}
for $t\ge 0$, and $\hat{S}(t)=0$ for $t < 0$.

\begin{figure}
\includegraphics[width=.8\columnwidth]{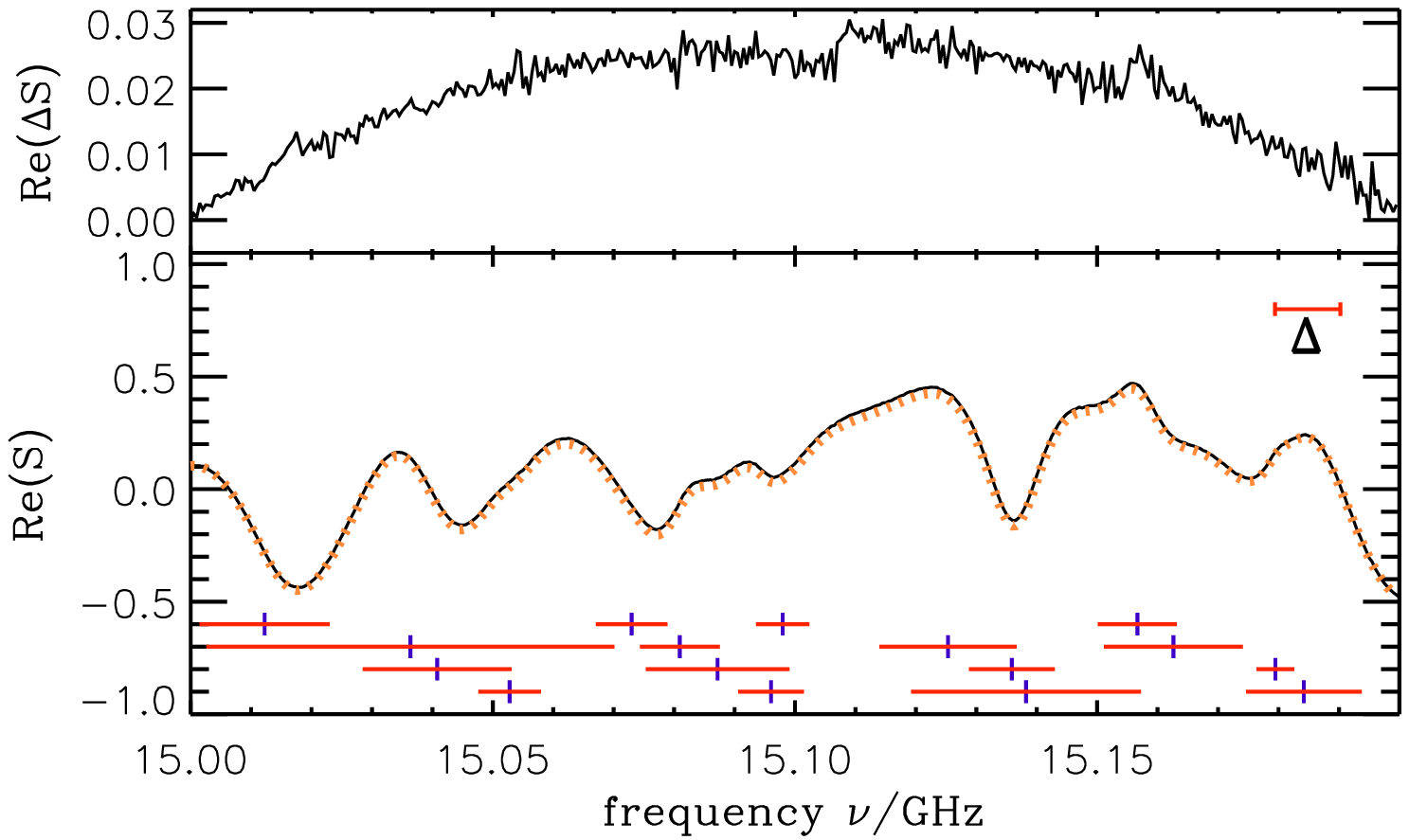}
\includegraphics[width=.8\columnwidth]{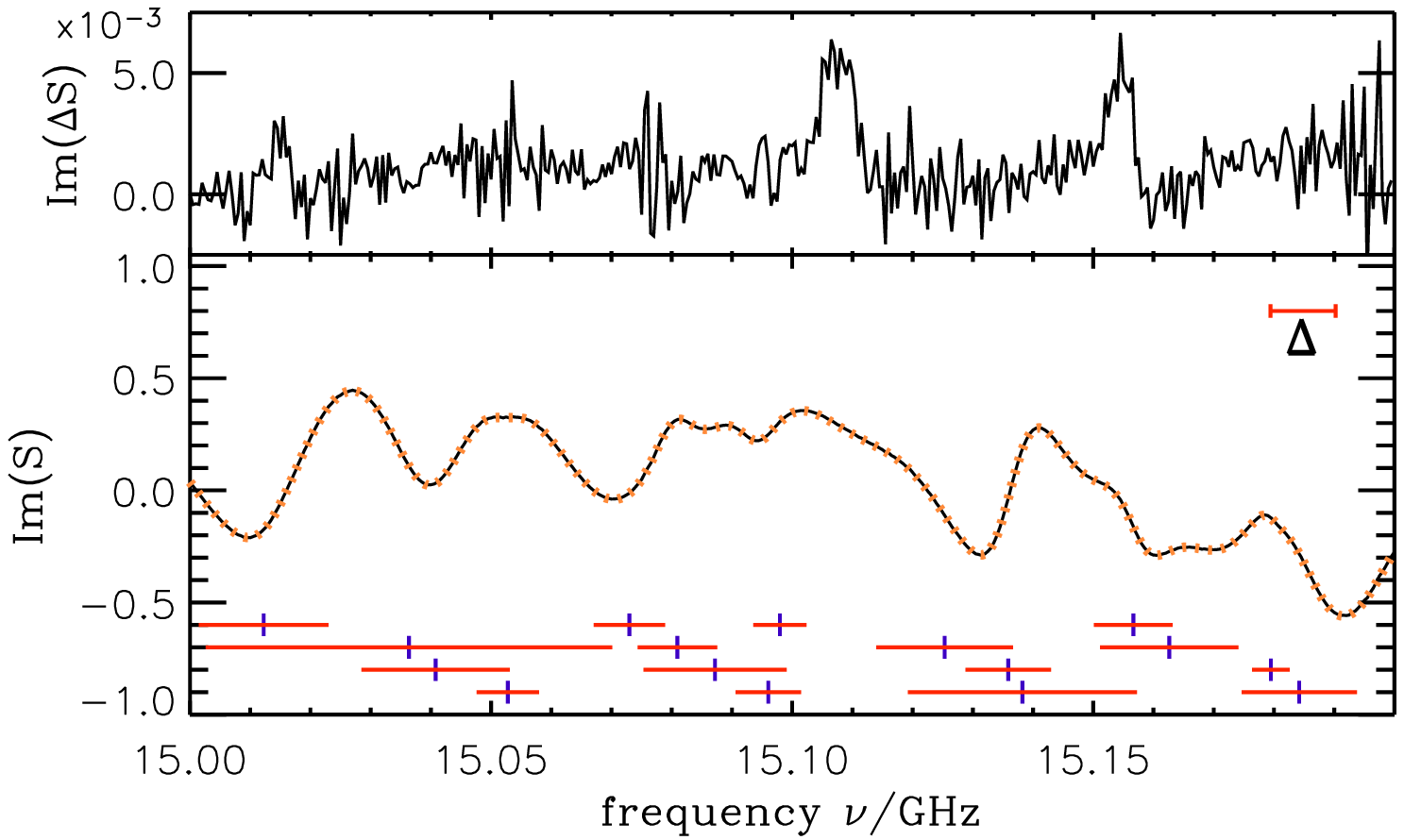}
\caption{\label{fig:Spectra} (Color online)
Real and imaginary part of a part of the spectrum in the regime of overlapping resonances. Additionally the reconstructed spectrum using the resonances identified by the harmonic inversion is shown in dashed line. The vertical lines indicate the positions of the resonances and the horizontal lines the corresponding widths at half maximum. On top of each figure the difference between the measured spectrum and the calculated one is shown. The mean level spacing $\Delta$ is marked by a horizontal bar.}
\end{figure}

Fig.~\ref{fig:Time} shows an example for $|\hat{S}(t)|^2$ obtained from the Fourier transform of the part of the spectrum shown in the inset. For the HI only the short time domain has been used, since the long time tail of the signal does contain noise only. The cut-off time is marked in Fig.~\ref{fig:Time} by a dotted vertical line. The number $N$ of data points has been of the order of a few hundred up to about 1000 points depending on the frequency range. A matrix of rank $N/2$ is created in the HI procedure \cite{mai99,wie08} and its eigenvalues and eigenvectors are calculated, yielding $N/2$ complex eigenfrequencies $\nu_n - i \gamma_n$ and their residua $a_n$. This number usually exceeds by far the real number of resonances in this frequency regime, which is approximately given by the Weyl formula
\begin{equation}
n(\nu)=\frac{\pi A}{c^2} \nu^2+\frac{S}{2c}\nu+K,
\end{equation}
where $A$ is the area, $S$ the circumference, $K$ is a constant of order one, and $c$ the speed of light. Weyl's formula counts the mean number of resonances below the frequency $\nu$ for closed cavities. In open systems the Weyl formula strictly speaking does not apply, but in the present situation the error should be small. Wall absorption essentially introduces an imaginary part for the resonances, and has only a minor influence on the density of states. Only the presence of the antenna may lead to the loss of a single resonance due to resonance trapping \cite{per00}.
To distinguish real from spurious resonances a number of criteria have been applied: (i) Resonances close to the border frequencies of the Fourier transform window have not been taken into account to avoid boundary effects. (ii) The HI has been performed twice, first for $\hat{S}(t)$, and additionally for $\hat{S}(t-t_0)$ corresponding to a shift of $\hat{S}(t)$ by one data point. Only resonances showing up to be stable with respect to this procedure have been taken. (iii) Resonances with a width close to the frequency spacing $\delta \nu$ between neighboring data points have been removed. (iv) Resonances with a depth $|a_n|/\gamma_n$ exceeding the noise level by less than a factor of two have been removed. The number of resonances surviving all checks amounted to about 80-85\% of the value expected from the Weyl formula.

In Fig.~\ref{fig:Spectra} the real and imaginary part of the evaluated spectrum is shown for a small frequency range in the regime of overlapping resonances. The solid line corresponds to the measured spectrum. The extracted resonances are shown as vertical lines and their widths at half maximum by horizontal lines. From the extracted resonances we reconstructed the spectrum in terms of a sum of Lorentzians (see Eq.~(\ref{eq:SumOfLorentzians})). The reconstructed spectrum was adjusted by removing an additional linear dependence, such that experimental and reconstructed spectrum coincide at both boundaries of the window. The reconstructed spectrum is plotted with a dashed line in Fig.~\ref{fig:Spectra}. On top the difference between measured and reconstructed spectrum is shown, both for the real and imaginary part. The main cause for the deviations are fluctuations in the experimental baseline. Since the HI is able only to treat superpositions of Lorentzians it cannot account for fluctuations of the baseline properly. But the local structure is in perfect agreement with the measured spectrum. For the frequency range shown in Fig.~\ref{fig:Spectra} Weyl's formula gives 18.37 resonances. We found 16 resonances indicating that we lost two to three resonances, probably due to too small amplitudes. Actually the difference between original and reconstructed spectrum suggests that there are two other resonances at 15.11 and 15.16\,GHz. They have not been taken into account due to the noise cut-off criterion applied.

\begin{figure}
\includegraphics[width=.8\columnwidth]{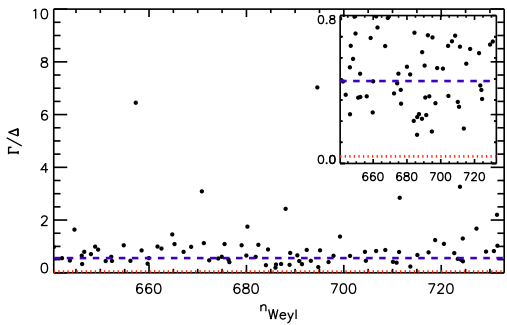}
\caption{\label{fig:Width} (Color online)
Normalized widths $\Gamma/\Delta$ for 14.7 to 15.7\,GHz with $T_t=7.00$ and $T_a=0.989$. The horizontal dashed line corresponds to the width coming from the total absorption ($\Gamma_t = T_t/(4\pi)$), whereas the dotted line (only visible in the inset) corresponds to the constant part of the absorption ($\Gamma_{\rm const}=T_{\rm const}/(4\pi))$. The inset is an enlargement of the small width region.}
\end{figure}

To compare our results from an electromagnetic resonator with the quantum-mechanical predictions, we mapped the experimental eigenfrequencies $\nu_n$ and their widths $\gamma_n$ to their quantum-mechanical counterparts by $E_n=\nu_n^2$ and $\Gamma_n=\nu_n\gamma_n$ \cite{kuh05b}. This procedure is justified as long as the linewidths $\gamma$ are small compared to the frequencies $\nu$, which was the case in the experiment ($\gamma/\nu \approx 10^{-3}$). In the effective Hamiltonian approach the quantum-mechanical reflection spectrum $S(E)$ can be described by
\begin{equation}
  \label{eq:Smatrix}
  S(E)=1-iW^\dagger\frac{1}{E-H_{\rm{eff}}}W,
\end{equation}
where $W$ contains the information on the coupling of the antenna to the eigenvalues of the closed system \cite{guh98}. The effective Hamiltonian is given by
\begin{equation}
  \label{eq:Heff}
  H_{\rm{eff}}=H_0-iWW^\dagger.
\end{equation}
Width distributions for a random matrix system with multiple coupled channels have been calculated by Sommers, Fyodorov and Titov \cite{som99}, reducing for the single channel case to
\begin{equation}
\label{eq:WidthDist}
P(y) = \frac{1}{4} \frac{\partial^2}{\partial y^2} \int_{-1}^1 {\rm{d}}\lambda \,\,\, (1-\lambda^2)e^{-2\pi \lambda y} F(\lambda,y)
\end{equation}
with
\begin{eqnarray}\label{Eq:PWidthF}
F(\lambda,y)&=&(g-\lambda)\int_g^\infty {\rm{d}}p_1 \frac{e^{\pi y p_1}}{(\lambda-p_1)^2\sqrt{(p_1^2-1)(p_1-g)} } \nonumber \\
&\times& \int_1^g {\rm{d}}p_2
\frac{(p_1-p_2)e^{\pi y p_2}}{(\lambda-p_2)^2\sqrt{(p_2^2-1)(g-p_2)}}
\end{eqnarray}
and $g=2/T_a-1$. Here $y=-\frac{\Gamma}{\Delta}$ is negative. In case of perfect coupling the tail is algebraically decaying as $1/(4\pi y^2)$.

\begin{figure}
\includegraphics[width=.8\columnwidth]{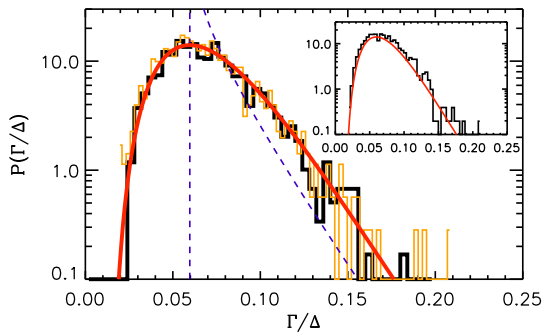}
\caption{\label{fig:PWidth1} (Color online)
Normalized width distribution in the frequency range 4 to 5\,GHz with $T_a=0.116$ and $T_t=0.75$. The black histogram was obtained from the harmonic inversion, the light one from a Lorentzian fitting procedure of the isolated resonances.
The dashed curve corresponds to the theoretical prediction from Eq.~(\ref{eq:WidthDist}) shifted by a constant off-set $\Gamma_t=T_t/(4\pi)$ to account for absorption in the walls. The solid line in addition takes into account the fluctuations using $T_{\rm fluc}=0.6$ and $N_c=10$ (see the text for details). The inset shows a comparison between a random matrix simulation and the theoretical prediction using the same values for $T_a$, $T_t$, $T_{\rm fluc}$, and $N_c$ as for the experimental data.}
\end{figure}

\begin{figure}
\includegraphics[width=.8\columnwidth]{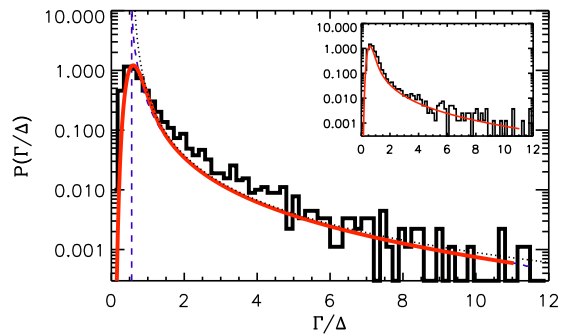}
\caption{\label{fig:PWidth3}
Same as Fig.~\ref{fig:PWidth1}, but for the frequency range 14.7 to 15.7\,GHz with $T_a=0.989$ and $T_t=7.0$. Additionally the asymptotics of the theoretical expectation for perfect coupling is plotted as a dotted curve (see text). For the simulations and the  theoretical curve $T_{\rm fluc}=6.4$ and $N_c=20$ have been used.}
\end{figure}

Let us now start with the discussion of the experimental results on resonance widths. For the part of the spectrum shown in Fig.~\ref{fig:Spectra} the resonance width is by far larger than the mean level spacing. This can be better seen in Fig.~\ref{fig:Width} where the widths of the resonances are plotted for the frequency range 14.7 to 15.7\,GHz in a regime of nearly perfect coupling ($T_a=0.989$) and large absorption ($T_t=7.0$) as a function of the mean level number calculated from the Weyl formula. All widths have been normalized to the mean level spacing $\Delta$. The horizontal dashed line corresponds to the contribution to the width from the total absorption, $\Gamma_t/\Delta = T_t/(4\pi)$ \cite{sav06} (in Ref.~\cite{sav06} there is a factor of $2\pi$ in the denominator, resulting from a differing definition of $\Gamma$).
A constant total absorption just shifts the theoretical curve. This has been considered in Figs.~\ref{fig:PWidth1} and \ref{fig:PWidth3} showing the experimental results in the regimes of weak and strong resonance overlap, respectively. The dashed lines correspond to the theoretical distribution (\ref{Eq:PWidthF}) shifted by $\Gamma_t/\Delta$. There are clear deviations for smaller widths, but the decay of the tails are in good agreement with the theoretical description, in particular for Fig.~\ref{fig:PWidth3}. The deviations are caused by the fact that there is a finite number of fictitious  channels $N_c$. Instead of using the exact formulas of Ref.~\cite{som99} for an arbitrary number of channels, which lead to serious stability problems in the numerical integrations, we folded the exact distribution (\ref{Eq:PWidthF}) with a $\chi^2$ distribution with $N_c$ degrees of freedom and a first moment of $\Gamma_{\rm fluc}=T_{\rm fluc}/(4\pi)$. The values of $N_c$ and $T_{\rm fluc}$ were chosen by minimizing the deviations from the experimental distributions. The insets in Figs.~\ref{fig:PWidth1} and \ref{fig:PWidth3} show the result of a random matrix numerical simulation, using expression (\ref{eq:Smatrix}) for the scattering matrix. The same number of eigenvalues has been taken as for the experimental histograms. The solid line corresponds to theory using the folding approximation.

Fig.~\ref{fig:PWidth1} represents the regime of isolated resonances. Here a standard fit procedure in terms of a superposition of Lorentzians was still possible. The resulting distribution of line widths shown as light line in Fig.~\ref{fig:PWidth1} is in agreement with the result from the HI. A perfect agreement between experiment and theory is obtained, but it is essential to take into account a coupling  to additional channels. Coupling strength $T_a$, and wall absorption $T_t$  have been obtained independently, as described above. Only the splitting of $T_t$ into a constant and a fluctuating part has been adjusted, yielding $T_{\rm fluc}=0.6$, $N_c=10$, $T_{\rm const}=0.15$.

Fig.~\ref{fig:PWidth3} shows the result in the regime of nearly perfect coupling, where there is a strong resonance overlap, too large to allow a direct fit.  Here the full potential of the HI technique becomes evident. We find a good coincidence between experiment and theory in the tail up to a width $\Gamma/\Delta $ = 12, already without considering the fluctuating part of the absorption, but if it is taken into account ($T_{\rm fluc}=6.4$, $N_c=20$, $T_{\rm const}=0.6$), an agreement over three orders of magnitude is found. In the moment the physical interpretation of the additional channels is not completely clear. If one tries to describe wall absorption by coupling a finite number of fictitious channels to top and bottom plate, or, alternatively, to the mantle \cite{sch03a}, the resulting channel numbers are by far too large. They might be due to capacitances occurring at interfaces between the mantle and top and bottom plates. The constant contribution to the absorption $T_{\rm const}$ is of the order of magnitude expected from the resistivity of brass.

In preliminary studies of systems with strong absorption we had already been able to analyze resonances up to the limit of $\Gamma/\Delta \approx 30$. This opens a wide variety of possibilities to study the properties of the poles of scattering matrices in the complex plane \cite{fyo97b}. The verification of the prediction by Sommers et al.\ \cite{som99} on the distribution of resonances is a first illustration in this respect.

We would like to thank F. Mortessagne, Nice, for fruitful comments.
The research was funded by the DFG via an individual grant and the Forschergruppe 760 `Scattering Systems with Complex Dynamics'.


\begin{thebibliography}{21}
\expandafter\ifx\csname natexlab\endcsname\relax\def\natexlab#1{#1}\fi
\expandafter\ifx\csname bibnamefont\endcsname\relax
  \def\bibnamefont#1{#1}\fi
\expandafter\ifx\csname bibfnamefont\endcsname\relax
  \def\bibfnamefont#1{#1}\fi
\expandafter\ifx\csname citenamefont\endcsname\relax
  \def\citenamefont#1{#1}\fi
\expandafter\ifx\csname url\endcsname\relax
  \def\url#1{\texttt{#1}}\fi
\expandafter\ifx\csname urlprefix\endcsname\relax\def\urlprefix{URL }\fi
\providecommand{\bibinfo}[2]{#2}
\providecommand{\eprint}[2][]{\url{#2}}

\bibitem[{\citenamefont{Kuhl et~al.}(2005{\natexlab{a}})\citenamefont{Kuhl,
  St\"{o}ckmann, and Weaver}}]{kuh05b}
\bibinfo{author}{\bibfnamefont{U.}~\bibnamefont{Kuhl}},
  \bibinfo{author}{\bibfnamefont{H.-J.} \bibnamefont{St\"{o}ckmann}},
  \bibnamefont{and} \bibinfo{author}{\bibfnamefont{R.}~\bibnamefont{Weaver}},
  \bibinfo{journal}{J. Phys. A} \textbf{\bibinfo{volume}{38}},
  \bibinfo{pages}{10433} (\bibinfo{year}{2005}{\natexlab{a}}).

\bibitem[{\citenamefont{Fyodorov et~al.}(2005)\citenamefont{Fyodorov, Savin,
  and Sommers}}]{fyo05a}
\bibinfo{author}{\bibfnamefont{Y.~V.} \bibnamefont{Fyodorov}},
  \bibinfo{author}{\bibfnamefont{D.~V.} \bibnamefont{Savin}}, \bibnamefont{and}
  \bibinfo{author}{\bibfnamefont{H.-J.} \bibnamefont{Sommers}},
  \bibinfo{journal}{J. Phys. A} \textbf{\bibinfo{volume}{38}},
  \bibinfo{pages}{10731} (\bibinfo{year}{2005}).

\bibitem[{\citenamefont{St\"{o}ckmann}(1999)}]{stoe99}
\bibinfo{author}{\bibfnamefont{H.-J.} \bibnamefont{St\"{o}ckmann}},
  \emph{\bibinfo{title}{Quantum Chaos - An Introduction}}
  (\bibinfo{publisher}{University Press}, \bibinfo{address}{Cambridge},
  \bibinfo{year}{1999}).

\bibitem[{\citenamefont{M\'ende{z-S\'a}nchez
  et~al.}(2003)\citenamefont{M\'ende{z-S\'a}nchez, Kuhl, Barth, Lewenkopf, and
  St\"{o}ckmann}}]{men03a}
\bibinfo{author}{\bibfnamefont{R.~A.} \bibnamefont{M\'ende{z-S\'a}nchez}},
  \bibinfo{author}{\bibfnamefont{U.}~\bibnamefont{Kuhl}},
  \bibinfo{author}{\bibfnamefont{M.}~\bibnamefont{Barth}},
  \bibinfo{author}{\bibfnamefont{C.~H.} \bibnamefont{Lewenkopf}},
  \bibnamefont{and} \bibinfo{author}{\bibfnamefont{H.-J.}
  \bibnamefont{St\"{o}ckmann}}, \bibinfo{journal}{Phys. Rev. Lett.}
  \textbf{\bibinfo{volume}{91}}, \bibinfo{pages}{174102}
  (\bibinfo{year}{2003}).

\bibitem[{\citenamefont{Kuhl et~al.}(2005{\natexlab{b}})\citenamefont{Kuhl,
  Mart\'{\i}nez-Mares, M\'endez-S\'anchez, and St\"{o}ckmann}}]{kuh05a}
\bibinfo{author}{\bibfnamefont{U.}~\bibnamefont{Kuhl}},
  \bibinfo{author}{\bibfnamefont{M.}~\bibnamefont{Mart\'{\i}nez-Mares}},
  \bibinfo{author}{\bibfnamefont{R.~A.} \bibnamefont{M\'endez-S\'anchez}},
  \bibnamefont{and} \bibinfo{author}{\bibfnamefont{H.-J.}
  \bibnamefont{St\"{o}ckmann}}, \bibinfo{journal}{Phys. Rev. Lett.}
  \textbf{\bibinfo{volume}{94}}, \bibinfo{pages}{144101}
  (\bibinfo{year}{2005}{\natexlab{b}}).

\bibitem[{\citenamefont{Hemmady et~al.}(2005)\citenamefont{Hemmady, Zheng, Ott,
  Antonsen, and Anlage}}]{hem05a}
\bibinfo{author}{\bibfnamefont{S.}~\bibnamefont{Hemmady}},
  \bibinfo{author}{\bibfnamefont{X.}~\bibnamefont{Zheng}},
  \bibinfo{author}{\bibfnamefont{E.}~\bibnamefont{Ott}},
  \bibinfo{author}{\bibfnamefont{T.~M.} \bibnamefont{Antonsen}},
  \bibnamefont{and} \bibinfo{author}{\bibfnamefont{S.~M.}
  \bibnamefont{Anlage}}, \bibinfo{journal}{Phys. Rev. Lett.}
  \textbf{\bibinfo{volume}{94}}, \bibinfo{pages}{014102}
  (\bibinfo{year}{2005}).

\bibitem[{\citenamefont{Barth\'elemy et~al.}(2005)\citenamefont{Barth\'elemy,
  Legrand, and Mortessagne}}]{bar05b}
\bibinfo{author}{\bibfnamefont{J.}~\bibnamefont{Barth\'elemy}},
  \bibinfo{author}{\bibfnamefont{O.}~\bibnamefont{Legrand}}, \bibnamefont{and}
  \bibinfo{author}{\bibfnamefont{F.}~\bibnamefont{Mortessagne}},
  \bibinfo{journal}{Europhys. Lett.} \textbf{\bibinfo{volume}{70}},
  \bibinfo{pages}{162} (\bibinfo{year}{2005}).

\bibitem[{\citenamefont{Lobkis et~al.}(2003)\citenamefont{Lobkis, Rozhkov, and
  Weaver}}]{lob03a}
\bibinfo{author}{\bibfnamefont{O.~I.} \bibnamefont{Lobkis}},
  \bibinfo{author}{\bibfnamefont{I.~S.} \bibnamefont{Rozhkov}},
  \bibnamefont{and} \bibinfo{author}{\bibfnamefont{R.~L.}
  \bibnamefont{Weaver}}, \bibinfo{journal}{Phys. Rev. Lett.}
  \textbf{\bibinfo{volume}{91}}, \bibinfo{pages}{194101}
  (\bibinfo{year}{2003}).

\bibitem[{\citenamefont{Akis et~al.}(1997)\citenamefont{Akis, Ferry, and
  Bird}}]{aki97}
\bibinfo{author}{\bibfnamefont{R.}~\bibnamefont{Akis}},
  \bibinfo{author}{\bibfnamefont{D.~K.} \bibnamefont{Ferry}}, \bibnamefont{and}
  \bibinfo{author}{\bibfnamefont{J.~P.} \bibnamefont{Bird}},
  \bibinfo{journal}{Phys. Rev. Lett.} \textbf{\bibinfo{volume}{79}},
  \bibinfo{pages}{123} (\bibinfo{year}{1997}).

\bibitem[{\citenamefont{Alt et~al.}(1995)\citenamefont{Alt, Gr\"{a}f, Harney,
  Hofferbert, Lengeler, Richter, Schardt, and Weidenm\"{u}ller}}]{alt95a}
\bibinfo{author}{\bibfnamefont{H.}~\bibnamefont{Alt}},
  \bibinfo{author}{\bibfnamefont{H.-D.} \bibnamefont{Gr\"{a}f}},
  \bibinfo{author}{\bibfnamefont{H.~L.} \bibnamefont{Harney}},
  \bibinfo{author}{\bibfnamefont{R.}~\bibnamefont{Hofferbert}},
  \bibinfo{author}{\bibfnamefont{H.}~\bibnamefont{Lengeler}},
  \bibinfo{author}{\bibfnamefont{A.}~\bibnamefont{Richter}},
  \bibinfo{author}{\bibfnamefont{P.}~\bibnamefont{Schardt}}, \bibnamefont{and}
  \bibinfo{author}{\bibfnamefont{H.~A.} \bibnamefont{Weidenm\"{u}ller}},
  \bibinfo{journal}{Phys. Rev. Lett.} \textbf{\bibinfo{volume}{74}},
  \bibinfo{pages}{62} (\bibinfo{year}{1995}).

\bibitem[{\citenamefont{Persson et~al.}(2000)\citenamefont{Persson, Rotter,
  St\"{o}ckmann, and Barth}}]{per00}
\bibinfo{author}{\bibfnamefont{E.}~\bibnamefont{Persson}},
  \bibinfo{author}{\bibfnamefont{I.}~\bibnamefont{Rotter}},
  \bibinfo{author}{\bibfnamefont{H.-J.} \bibnamefont{St\"{o}ckmann}},
  \bibnamefont{and} \bibinfo{author}{\bibfnamefont{M.}~\bibnamefont{Barth}},
  \bibinfo{journal}{Phys. Rev. Lett.} \textbf{\bibinfo{volume}{85}},
  \bibinfo{pages}{2478} (\bibinfo{year}{2000}).

\bibitem[{\citenamefont{Fyodorov et~al.}(1997)\citenamefont{Fyodorov,
  Khoruzhenko, and Sommers}}]{fyo97a}
\bibinfo{author}{\bibfnamefont{Y.~V.} \bibnamefont{Fyodorov}},
  \bibinfo{author}{\bibfnamefont{B.~A.} \bibnamefont{Khoruzhenko}},
  \bibnamefont{and} \bibinfo{author}{\bibfnamefont{H.-J.}
  \bibnamefont{Sommers}}, \bibinfo{journal}{Phys. Lett. A}
  \textbf{\bibinfo{volume}{226}}, \bibinfo{pages}{46} (\bibinfo{year}{1997}).

\bibitem[{\citenamefont{Fyodorov and Sommers}(1997)}]{fyo97b}
\bibinfo{author}{\bibfnamefont{Y.~V.} \bibnamefont{Fyodorov}} \bibnamefont{and}
  \bibinfo{author}{\bibfnamefont{H.-J.} \bibnamefont{Sommers}},
  \bibinfo{journal}{J. Math. Phys.} \textbf{\bibinfo{volume}{38}},
  \bibinfo{pages}{1918} (\bibinfo{year}{1997}).

\bibitem[{\citenamefont{Sommers et~al.}(1999)\citenamefont{Sommers, Fyodorov,
  and Titov}}]{som99}
\bibinfo{author}{\bibfnamefont{H.~J.} \bibnamefont{Sommers}},
  \bibinfo{author}{\bibfnamefont{Y.~V.} \bibnamefont{Fyodorov}},
  \bibnamefont{and} \bibinfo{author}{\bibfnamefont{M.}~\bibnamefont{Titov}},
  \bibinfo{journal}{J. Phys. A} \textbf{\bibinfo{volume}{32}},
  \bibinfo{pages}{L77} (\bibinfo{year}{1999}).

\bibitem[{\citenamefont{Wall and Neuhauser}(1995)}]{wal95}
\bibinfo{author}{\bibfnamefont{M.~R.} \bibnamefont{Wall}} \bibnamefont{and}
  \bibinfo{author}{\bibfnamefont{D.}~\bibnamefont{Neuhauser}},
  \bibinfo{journal}{J. Chem. Phys.} \textbf{\bibinfo{volume}{102}},
  \bibinfo{pages}{8011} (\bibinfo{year}{1995}).

\bibitem[{\citenamefont{Mandelshtam and Taylor}(1997)}]{man97b}
\bibinfo{author}{\bibfnamefont{V.~A.} \bibnamefont{Mandelshtam}}
  \bibnamefont{and} \bibinfo{author}{\bibfnamefont{H.~S.}
  \bibnamefont{Taylor}}, \bibinfo{journal}{Phys. Rev. Lett.}
  \textbf{\bibinfo{volume}{78}}, \bibinfo{pages}{3274} (\bibinfo{year}{1997}).

\bibitem[{\citenamefont{Main}(1999)}]{mai99}
\bibinfo{author}{\bibfnamefont{J.}~\bibnamefont{Main}}, \bibinfo{journal}{Phys.
  Rep.} \textbf{\bibinfo{volume}{316}}, \bibinfo{pages}{233}
  (\bibinfo{year}{1999}).

\bibitem[{\citenamefont{Wiersig and Main}(2008)}]{wie08}
\bibinfo{author}{\bibfnamefont{J.}~\bibnamefont{Wiersig}} \bibnamefont{and}
  \bibinfo{author}{\bibfnamefont{J.}~\bibnamefont{Main}},
  \bibinfo{journal}{Phys. Rev. E} \textbf{\bibinfo{volume}{77}},
  \bibinfo{pages}{036205} (\bibinfo{year}{2008}).

\bibitem[{\citenamefont{Sch\"{a}fer et~al.}(2003)\citenamefont{Sch\"{a}fer,
  Gorin, Seligman, and St\"{o}ckmann}}]{sch03a}
\bibinfo{author}{\bibfnamefont{R.}~\bibnamefont{Sch\"{a}fer}},
  \bibinfo{author}{\bibfnamefont{T.}~\bibnamefont{Gorin}},
  \bibinfo{author}{\bibfnamefont{T.~H.} \bibnamefont{Seligman}},
  \bibnamefont{and} \bibinfo{author}{\bibfnamefont{H.-J.}
  \bibnamefont{St\"{o}ckmann}}, \bibinfo{journal}{J. Phys. A}
  \textbf{\bibinfo{volume}{36}}, \bibinfo{pages}{3289} (\bibinfo{year}{2003}).

\bibitem[{\citenamefont{Savin et~al.}(2006)\citenamefont{Savin, Legrand, and
  Mortessagne}}]{sav06}
\bibinfo{author}{\bibfnamefont{D.~V.} \bibnamefont{Savin}},
  \bibinfo{author}{\bibfnamefont{O.}~\bibnamefont{Legrand}}, \bibnamefont{and}
  \bibinfo{author}{\bibfnamefont{F.}~\bibnamefont{Mortessagne}},
  \bibinfo{journal}{Europhys. Lett.} \textbf{\bibinfo{volume}{76}},
  \bibinfo{pages}{774} (\bibinfo{year}{2006}).

\bibitem[{\citenamefont{Guhr et~al.}(1998)\citenamefont{Guhr,
  M\"{u}ller-Groeling, and Weidenm\"{u}ller}}]{guh98}
\bibinfo{author}{\bibfnamefont{T.}~\bibnamefont{Guhr}},
  \bibinfo{author}{\bibfnamefont{A.}~\bibnamefont{M\"{u}ller-Groeling}},
  \bibnamefont{and} \bibinfo{author}{\bibfnamefont{H.~A.}
  \bibnamefont{Weidenm\"{u}ller}}, \bibinfo{journal}{Phys. Rep.}
  \textbf{\bibinfo{volume}{299}}, \bibinfo{pages}{189} (\bibinfo{year}{1998}).

\end{thebibliography}
\end{document}